\documentclass[12pt]{article}

\usepackage{amsmath,amssymb}
\usepackage{rcdcom}
\usepackage[graph]{rcdgraph}

\usepackage[cp1251]{inputenc}
\usepackage[T2A,OT1]{fontenc}
\usepackage[english]{babel}

\newtheorem{theorem}{Theorem}

\newtheorem{proposition}{Proposition}

\newtheorem{remark}{Remark}

\begin{document}
{\bf NEW PERIODIC SOLUTIONS FOR THREE OR FOUR IDENTICAL VORTICES ON A PLANE AND A
SPHERE}\footnote{
Keywords and phrases: periodic solutions, choreographies, vortex dynamics,
reduction.

2000 Mathematics Subject Classification: 37N10, 76B47.

This work was supported in part by CRDF (RU-M1-2583-MO-04),
INTAS (04-80-7297), RFBR (04-205-264367) and NSh (136.2003.1).}

\begin{center}
{A.\,V.\,Borisov, I.\,S.\,Mamaev, A.\,A.\,Kilin}\\
Institute of Computer Sciences\\
Udmurt State University\\
426034  Izhevsk, Russia\\
E-mails: borisov@rcd.ru, mamaev@rcd.ru, aka@rcd.ru\\
\end{center}

\begin{abstract}
In this paper we describe new classes of periodic
solutions for point vortices on a plane and a sphere. They correspond to
similar solutions (so-called choreographies) in celestial mechanics.
\end{abstract}

\section{Equations of motion and first integrals for vortices on a
plane}
For~$n$ point vortices with Cartesian coordinates~$(x_i,y_i)$ and
intensities~$\Gamma_i$, the Hamiltonian equations of the motion are
\begin{equation}
\label{bmk-eq-1} \Gamma_i \dot x_i=\pt{\mathcal H}{y_i},\quad \Gamma_i \dot
y_i=-\pt{\mathcal H}{x_i},\quad 1\le i\le n,
\end{equation}
where the Hamiltonian is
\begin{equation}
\label{bmk-eq-2}
{\mathcal H}=-\frac{1}{4\pi}\sum_{i<j}^n{}\Gamma_i\Gamma_j\ln|\bs r_i-\bs
r_j|^2,\quad\bs r_i=(x_i,y_i).
\end{equation}
For these equations, the Poisson bracket
is~$\{x_i,\,y_j\}=\Gamma_i^{-1}\delta_{ij}$.

The system \eqref{bmk-eq-1} has three first integrals
\begin{equation}
\label{bmk-eq-3}
Q=\sum_{i=1}^n\Gamma_i x_i,\quad P=\sum_{i=1}^n \Gamma_i y_i,\quad
I=\sum_{i=1}^n \Gamma_i(x_i^2+y_i^2),
\end{equation}
which are not involutive:
\begin{equation}
\label{bmk-eq-4}
\{Q,P\}=\sum_{i=1}^N\Gamma_i,\quad \{P,I\}=-2Q,\quad \{Q,I\}=2P.
\end{equation}

However, two involutive integrals always exist, for example,~$Q^2+P^2$ and
$I$. With the help of these integrals, we can reduce the system by two
degrees of freedom.

Thus, in the case of three vortices the reduced system has one degree of
freedom and it is integrable (Gr\"obli, 1877~\cite{grebli}; Poincar\'{e}, 1893~\cite{new_1}).

The four-vortex problem is reduced to a system with two degrees of
freedom. This system is not integrable (S.\,Ziglin, 1979 \cite{bmk-7}).

\section{Equations of motion and first integrals for vortices on a
sphere~$\mathbb{S}^2$}

For~$n$ point vortices moving on a sphere~$\mathbb{S}^2$, with spherical
coordinates~$(\theta_i,\varphi_i)$ and intensities~$\Gamma_i$, the
Hamiltonian equations can be written as (Bogomolov, 1977~\cite{bogomol})
\begin{equation}
\label{bmk-eq-6} \dot{\theta}_i=\{\theta_i,{\mathcal H}\},\quad
\dot{\varphi}_i=\{\varphi_i,{\mathcal H}\},\quad i=1,\ldots,n,
\end{equation}
with the Poisson
bracket~$\{\varphi_i,\cos\theta_k\}=\frac{\delta_{ik}}{R^2\Gamma_i}$ and
the Hamiltonian
\begin{equation}
\label{bmk-eq-8} {\mathcal
H}=-\frac1{4\pi}\sum_{i<k}^n\Gamma_i\Gamma_k\ln\left(4R^2\sin^2\frac{\gamma_{ik}}2\right).
\end{equation}
Here, $R$ is the sphere's radius, and $\gamma_{ik}$ is the angle between the
vectors from the center of the sphere to the point vortices~$i$ and~$k$ so
that
$$
\cos\gamma_{ik}=\cos\theta_i\cos\theta_k+\sin\theta_i\sin\theta_k\cos(\varphi_i-\varphi_k).
$$

Beside the Hamiltonian, the equations~(\ref{bmk-eq-6}) have three
independent non-involutive integrals
\begin{equation}
\label{bmk-eq-9}
F_1=R\sum_{i=1}^n{}\Gamma_i\sin\theta_i\cos\varphi_i,\quad
F_2=R\sum_{i=1}^n{}\Gamma_i\sin\theta_i\sin\varphi_i,\quad
F_3=R\sum_{i=1}^n{}\Gamma_i\cos\theta_i.
\end{equation}

The vector~$\bs F=\sum\Gamma_i \bs r_i$ ($\bs r_i$ are the radius-vectors of the vortices)
 with components~$(F_1,\,F_2,\,F_3)$  is called the
\emph{moment or center of vorticity\/}. The first integrals commute in the following
manner:~$\{F_i,F_j\}=\frac1{R}\varepsilon_{ijk}F_k$. Similar to the case
of a plane, we can reduce the system by two degrees of freedom, using
involutive integrals, for example,~$F_3,\,{|\bs F|}^2$.

So, in the case of three vortices, we have a completely integrable system
(P.\,New\-ton and R.\,Kidambi,~1998~\cite{kidambi}, 2000~\cite{bmk-2}; Borisov and Lebedev,
1998~\cite{bmlit5}). The four-vortex problem is reduced to a system with
two degrees of freedom and, generally, is not integrable (D.\,A. and
A.\,A.\,Bagrets~\cite{bagrets}).

\section{Reduction of a system of vortices on a plane}

To find new periodic solutions and apply the numerical analysis, we will
do the most symmetric reduction of the system by two degrees of freedom.
We present here one of the possible ways to do such a reduction, based on
new equations of motion for the mutual variables.

Effective reduction of a four-vortex system was done by
Khanin~\cite{bmk-4} (for vortices of the same sign) and by Aref and
Pomphrey~\cite{aref} (for equal vortices). The generalized Jacobi
reduction by one degree of freedom in the point vortex dynamics was
described in~\cite{bmk-8}.

In our study, we used the universal algebraic method of effective
reduction (for an arbitrary number of vortices). It is based on
\emph{mutual variables representation} of the motion equations of a system
of point vortices~\cite{bmk-9}, \cite{new_2}, \cite{new_3}.

As mutual variables, we introduce the following quantities.
Squared distances between the pairs
of vortices~$M_{ij}$ and oriented areas of triangles~$\Delta_{ijk}$,
$$
M_{ij}=(x_i-x_j)^2+(y_i-y_j)^2,\quad\Delta_{ijk}=(\bs r_j-\bs
r_k)\wedge(\bs r_k-\bs r_i).
$$
These variables are due to E.\,Laura.

We put~$\Gamma_i=\Gamma_j=1$, $P=Q=0$. Then the momentum
integral~$I$~\eqref{bmk-eq-3} can be written as
\begin{equation}
\label{D1}
I=\frac1n\sum\limits_{i<j}^nM_{ij},
\end{equation}
where $n$ is the number of the vortices. Writing the vortices' coordinates
in the complex form~$z_k=x_k+iy_k$, we obtain the following representation
for them:
\begin{equation}
\label{D2}
z_k=\frac1n\sum\limits_{j=k}^n\sqrt{M_{kj}}e^{i\theta_{kj}},
\end{equation}
where $\theta_{kj}$ is the angle between the vector from the~$j$th vortex
to the~$k$th vortex and the positive direction of~$Ox$.

By direct calculation we can derive the following propositions, which
describe the behavior of the reduced system of three and four vortices~\cite{bmk-9}.

\begin{proposition}\label{pro1}
For three identical vortices, the evolution of the mutual distances
$($with fixed~$I=\const)$ is described by a Hamiltonian system with one
degree of freedom. In terms of the canonical variables~$(g,G)$, this
system can be written as
\begin{equation}
\label{D3}
\dot g=\frac{\partial{\mathcal H}}{\partial G},\quad
\dot G=-\frac{\partial{\mathcal H}}{\partial g},\quad
{\mathcal H}=-\frac1{4\pi}\ln M_{12}M_{13}M_{23},
\end{equation}
where $M_{12}=4\Bigl(\frac I2{-}G\Bigr)$, $M_{13}=8G-I+2\sqrt{12}
\sqrt{\Bigl(\frac I2{-}G\Bigr)G}\cos g$, $M_{23}=4\Bigl(\frac
I2{-}G\Bigr)-2\sqrt{12} \sqrt{\left(\frac I2-G\right)G}\cos g$.
\end{proposition}

\begin{proposition}\label{pro2}
For four identical vortices, the evolution of the mutual distances is
described by a Hamiltonian system with two degrees of freedom. In terms of
the canonical variables~$(g,G,h,H)$, this system is
\begin{gather}
\label{D4}
\dot g=\frac{\partial{\mathcal H}}{\partial G},\quad
\dot G=-\frac{\partial{\mathcal H}}{\partial g},\quad
\dot h=\frac{\partial{\mathcal H}}{\partial H},\quad
\dot H=-\frac{\partial{\mathcal H}}{\partial h},\\
{\mathcal H}=-\frac1{4\pi}\ln M_{12}M_{13}M_{14}M_{23}M_{24}M_{34},
\notag
\end{gather}
where
$$
\begin{aligned}
M_{12}&=I{-}G{+}2\sqrt{(I{-}H)(I{-}G)}\cos h,&\quad
M_{34}&=I{-}G{-}2\sqrt{(I{-}H)(I{-}G)}\cos h,\\
M_{13}&=I{+}G{+}2\sqrt{(I{-}H)G}\cos(h{+}g),&\quad
M_{24}&=I{+}G{-}H{-}2\sqrt{(I{-}H)G}\cos(h{+}g),\\
M_{14}&=H{+}2\sqrt{(H{-}G)G}\cos g,&\quad
M_{23}&=H{-}2\sqrt{(I{-}G)G}\cos g.
\end{aligned}
$$
\end{proposition}

\begin{remark}
The above canonical variables naturally follow from the Lie-algebraic
representation of the equations of motion.
\end{remark}

\section{Absolute motion --- quadratures and geometric interpretation.}
According to \eqref{D2}, when $M_{ij}(t)$ are known, one needs to find the
angles~$\theta_{ij}(t)$ in order to determine the coordinates of the
vortices. Clearly, only one of these angles is independent (let it
be~$\theta_{12}$), the remaining are obtained using the
relations
\begin{equation} \label{star1}
\theta_{ij}+\theta_{ik}=\arccos\left(\frac{M_{jk}-M_{ij}-M_{ik}}{2\sqrt{2M_{ij}M_{ik}}}\right),
\quad i\ne j,\;k\ne i.
\end{equation}
The evolution of~$\theta_{ij}$ is found by integrating the first order equation:
\begin{equation}
\label{star}
4\pi\dot\theta_{ij}=\frac{2}{M_{ij}}\sum_{k=1}^{n}\Gamma_k+\sum_{k\ne
i,j}^{n}\Gamma_k\left(\frac{1}{M_{ik}}+\frac{1}{M_{jk}}\right)-
\frac{1}{M_{ij}}\sum_{k\ne i,j}^{n}\Gamma_k\left(\frac{M_{jk}}{M_{ik}}+\frac{M_{ik}}{M_{jk}}\right).
\end{equation}

There is an interesting geometric interpretation of the absolute motion of
the periodic solutions of the reduced systems \eqref{D3}, \eqref{D4}.

\begin{figure}
\begin{center}
\includegraphics{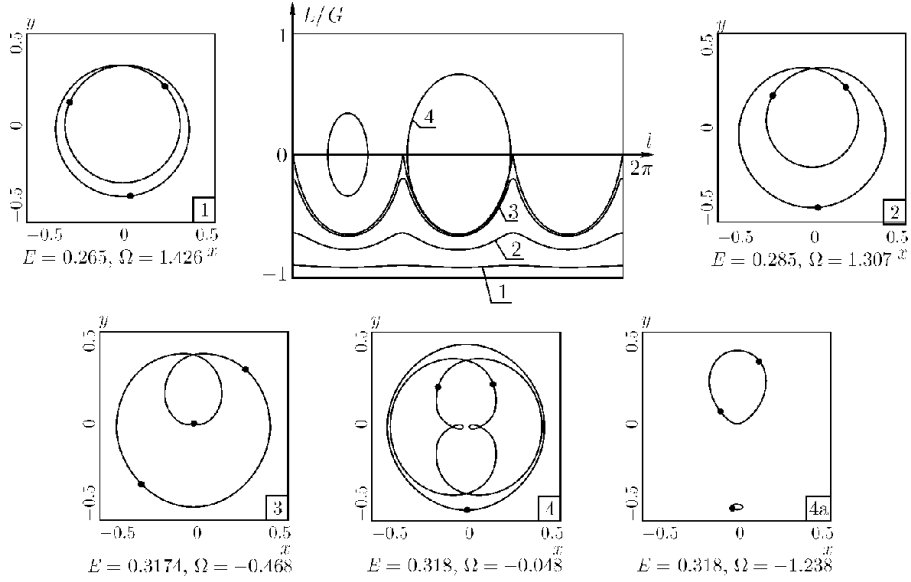}
\caption{The phase portrait of the reduced
system in the three-vortex problem, and the relative choreographies,
corresponding to different phase orbits in the portrait.}
\end{center}
\end{figure}

\begin{proposition}\label{pro3}
Let $\gamma(t)$ be a periodic solution $($of period~$T)$ of the reduced
system. Then\/{\rm

$1^\circ$} there exists a frame of reference, rotating with the constant
angular velocity~$\Omega_a$ about the center of vorticity, where each
vortex moves along some closed curve~$\xi_i(t);$

{\rm$2^\circ$} the rotating velocity $\Omega_a$ is given by (accurate
to~$\frac{2\pi}T\frac pq$, $p,q\in\mathbb{Z})${\rm:
\begin{equation}
\label{Om1}
\Omega_a=\frac1T\int\limits_0^T\dot\theta_{12}(t)dt;\vspace{-2mm}
\end{equation}

$3^\circ$} if $\Omega_a$ and $\Omega_o=\frac{2\pi}T$ are commensurable
$($i.\,e.~$\frac{\Omega_a}{\Omega_o}=\frac pq$, $p,q\in\mathbb{Z}),$ then
in the fixed frame of reference the vortices also move along closed
curves\/{\rm;

$4^\circ$} if any of the curves~$\xi_i(t)$ transform into each other by
rotation about the center of vorticity by an angle, commensurable
to~$2\pi$, then there is a (rotating) frame of reference where the
corresponding vortices orbit one and the same curve.
\end{proposition}

PROOF

In the right hand side of~\eqref{star}, there is a periodic function of
period~$T$. We have a convergent Fourier expansion:\vspace{-2mm}
\begin{equation}
\label{wer2}
4\pi\dot\theta_{12}=\sum_{n\in\mathbb{Z}}a^{(n)}e^{i\frac{2\pi n}Tt}.
\end{equation}
Integrating~\eqref{wer2}, and using~\eqref{star1}, we find that all the
angles~$\theta_{ij}$ depend on~$t$ in the following way:
\begin{equation}
\label{wer3}
\theta_{ij}(t)=\Omega_{ij} t+g_{ij}(t),
\end{equation}
where $\Omega_{ij}=a^{(0)}+\frac{2\pi}{T}\frac{q_{ij}}{p_{ij}},\q
q_{ij},\,p_{ij}\in\mathbb{Z}$, while~$g_{ij}(t)=g_{ij}(t+T)$
are~$T$-periodic functions of time.

Substituting the above dependency into~\eqref{D2}, we find that the
positions of the vortices on the plane are given as follows: \vspace{-1mm}
\begin{equation}
\label{wer31}
z_k(t)=\frac{Q+iP}{\sum\Gamma_i}+u_k(t)e^{i\Omega t},\q
u_k(t)=u_k(t+T_1)\in\mathbb{C},\q
\Omega=a^{(0)}.
\end{equation}
It follows that {\it in the frame of reference, rotating about the center
of vorticity with the angular velocity~$\Omega$, all the vortices describe
analytic closed curves, given by functions~$u_k(t)\in\mathbb{C}$}.

The proofs of $2^\circ$, $3^\circ$, and~$4^\circ$, using the
relations~\eqref{wer2}--\eqref{wer31}, are obvious. $\blacksquare$

\begin{remark}
Without any modification, Proposition \ref{pro3} is generalized to the
case of~$n$ vortices if we suppose that~$\gamma(t)$ is a periodic solution
of a reduced system with~$2n-2$ degrees of freedom to the~$n$-vortex
problem.
\end{remark}
\section{Particular solutions and stationary configurations for~$n$ equal
vortices}

This section presents most well-known particular solutions for point
vortices on a plane and on a sphere.

\paragraph*{\emph{The case of}~$\mathbb{R}^2$:}
\begin{enumerate}
\item collinear configurations;
\item polygonal configurations (in particular, polygonal configurations,
embedded into each other). An~$n$-polygon (Thomson
configuration~\cite{Tomson}) is stable if $n\le 7$, and unstable if~$n>7$ (the final
result on the stability of the heptagon was obtained recently in~\cite{Kurakin});
\item non-symmetrical stationary configuration (for~$n\ge 8$)~\cite{glass}.
\end{enumerate}

\paragraph*{\emph{The case of}~$\mathbb{S}^2$:}
\begin{enumerate}
\item Collinear configuration.

The roots~$\theta_i$ that define a collinear configuration, rotating with
velocity~$\Omega$, can be found as equilibriums of a system of~$N$
particles on a circle with the Hamiltonian \vspace{-1mm}
\begin{equation}\label{bmk-eq-10a}
{\mathcal H}=\frac12\sum_{k=1}^n p_k^2+4\pi
R^2\Omega\sum_{k=1}^N\cos\theta_k+\frac12\sum_{k,\,i=1}^N
{}'\ln\left|\sin\left(\frac{\theta_k-\theta_i}2\right)\right|.
\end{equation}

This system was obtained and studied by A.\,V.\,Borisov and V.\,V.\,Kozlov
(1999 \cite{bk-dan})~--- a special case with~$\Omega=0$ is called the Dyson system. It
is proved that the system (\ref{bmk-eq-10a}) is not integrable, but a
\emph{quasi-integral} exists, which gives a very good approximation of the
behavior of systems at low energy.
\item Analogs of Thomson configurations~\cite{bogomol-2}.

In these configurations, the vortices are positioned at the same
latitude~$\theta_0$ in the vertices of a regular~$n$-polygon and rotate
about its center with the angular velocity \vspace{-2mm}
$$
\omega=\frac{\Gamma(N-1)}{4\pi R^2}\frac{\ctg\theta_0}{\sin\theta_0}.
$$
\end{enumerate}

\section{Choreographies}

Recently, a new class of periodic solutions to the classical~$n$-body
problem has been obtained in celestial mechanics (C.\,Moore,
R.\,Montgomery, A.\,Chenciner, C.\,Sim\'o, and others). For example, they
found a remarkable stable periodic orbit in the~3-body problem~--- namely,
a \emph{figure-of-eight\/}~\cite{bmlit7}. In this solution, three bodies
follow each other along the same curve.

A long and ingenious analytical proof was done using the variational method based on
the least action principle. These solutions where all particles move along the same
curve are called \emph{choreographies}~\cite{new_4}.

Choreographies can be:
\begin{itemize}
\item [a)] \emph{absolute} (in the fixed frame of reference) or \emph{relative} (in
rotating frames of reference);
\item [b)] \emph{simple} (a single closed curve) or \emph{complex} (more
than one closed curve)
\end{itemize}

Similar solutions exist for point vortex dynamics on a plane and on a
sphere~\cite{bmlit8}. However, the least action principle cannot be
applied to an analytical proof in this case.

Using Proposition \ref{pro3} and the known analytical
solutions~\cite{grebli}, one can prove the following theorem for
three-vortex systems:
\begin{theorem}\label{teo1}
If, for the motion of three vortices of equal intensity, the constants of
motion,~$I$ and~$\mathcal H$, satisfy the inequalities
\begin{equation}
\label{D5} -\ln 3<\frac{4\pi}3{\mathcal H}+\ln I<\ln 2,
\end{equation}
then such motion is a simple relative choreography~(see
Fig.~\ref{plane.eps}).
\end{theorem}

\begin{remark}
Note also that various properties of the motion of three vortices are
discussed in~\cite{new_2} and~\cite{new_3}. For
example, the paper~\cite{new_3} focuses on the study of stability of collinear
configurations. However, properties of the absolute motion have not
actually been studied.
\end{remark}

The four-vortex system has a remarkable symmetric analytical solution~---
Goryachev's solution~\cite{GoryachevSbornik} (see also ~\cite{aref1})
where the vortices form a parallelogram as they move on a plane.

\begin{figure}[!ht]
\begin{center}
\cfig<bb=0 0 101.0mm 35.2mm>{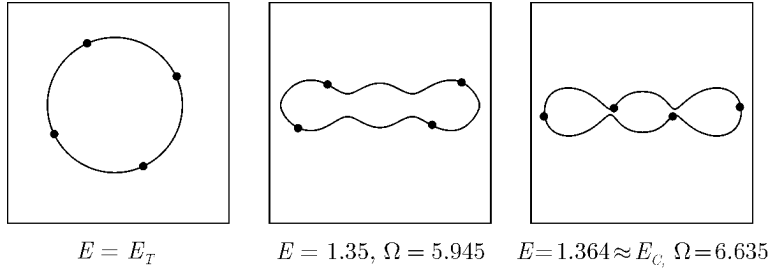}
\end{center}
\caption{Relative choreographies for Goryachev's solution}\label{4vortpar.eps}\
\end{figure}

\begin{theorem}\label{teo2}
If, for the motion of four vortices, the vortices $($of equal
intensity\/$)$ form a centrosymmetrical configuration $($a
parallelogram\/$)$, while the constants~$\mathcal H$ and $I$ satisfy the
inequalities\vspace{-2mm}
$$
-\ln2<\frac{2\pi}3{\mathcal H}+\ln I<-\ln\frac{144}5,
$$
then the motion is a simple relative choreography (see
Fig.~\ref{4vortpar.eps}).
\end{theorem}

For the complete proofs, see~\cite{bmlit8}.

Let us show that for the system of four identical vortices, there exist
choreographies, which do not coincide with Goryachev's solution. These are
\emph{non-analytical choreographies\/}. These solutions can be obtained by
variation of the Hamiltonian near Thomson's solution.

Indeed, near the Thomson configuration, the Hamiltonian can be written in
as \vspace{-2mm}
$$
{\mathcal H}={\mathcal H}_2+{\mathcal H}^*,\qquad \deg{\mathcal H}\ge 3,
$$
where ${\mathcal H}_2$ is a normal form\vspace{-2mm}
$$
{\mathcal H}_2=\frac1{4\pi}\left(3(x^2+P^2_x)+2\sqrt{2}(y^2+P^2_y)\right),
$$
and ${\mathcal H}^*$ is the perturbation function (a term of higher order).

On the level ${\mathcal H}_2=h_2=\const$, there are precisely two
nondegenerate periodic motions of the Hamiltonian system with the
Hamiltonian~${\mathcal H}_2$:
\begin{equation}\label{nchor1}
x=P_x=0,\qquad y^2+P_y^2=\frac{\pi}{\sqrt{2}}h_2,\vspace{-2mm}
\end{equation}
\begin{equation}\label{nchor2}
y=P_y=0,\qquad x^2+P_x^2=\frac{4\pi}3 h_2.
\end{equation}

\begin{figure}[!ht]
\begin{center}
\cfig<bb=0 0 102.6mm 37.3mm>{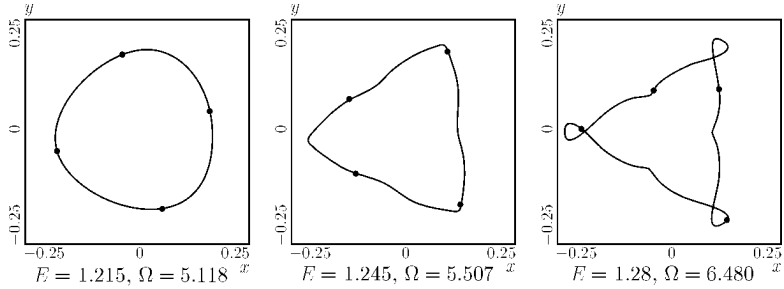}
\end{center}
\caption{Relative choreographies for the solution~(\ref{nchor2}).}\label{trapez.eps}
\end{figure}

According to the Lyapunov theorem, these periodic solutions are conserved
under the perturbation~${\mathcal H}^*$ and near Thomson's solution there
are two periodic solutions of the whole system~${\mathcal H}_2+{\mathcal
H}^*$:
\begin{itemize}
\item[] solution (\ref{nchor1}) gives Goryachev's solution, and
\item[] solution (\ref{nchor2}) gives a non-analytical relative choreography such
as shown in Fig.~\ref{trapez.eps}.
\end{itemize}

\begin{remark}
Here a solution is called non-analytical if it cannot be
expressed explicitly in terms of quadratures. As a rule, this means that
there are no explicit symmetries for these solutions, since all known symmetrical
solutions can usually be found in terms of quadratures.
\end{remark}

\bigskip

\section{Absolute choreographies}

Let us discuss in more detail, whether absolute choreographies exist in
the three- and four-vortex problems. According to the above, any relative
choreography, corresponding to a periodic solution (of period~$T$) of the
reduced system (see~\eqref{D3},~\eqref{D4}) closes in
time~$mT,\,m\in\mathbb{N}$. During this interval the vortices pass through
the same configurations~$m$ times. We denote the corresponding angular
velocities of the frames of reference by~$\Omega_m^{(k)}$.

For three-vortex choreographies (Theorem~\ref{teo1}), there exists a
(rotating) frame of reference, where a choreography closes in the smallest
possible time~$T$. For the case of four-vortex choreographies
(Theorem~\ref{teo2}), this time equals~$2T$. We denote and the corresponding
angular velocities by~$\Omega_1^{(0)}$ and~$\Omega_2^{(0)}$. The angular
velocities of other relative connected choreographies are now given by the
following relations.

For three vortices:
$$
\Omega_m^{(k)}(E)=\Omega_1^{(0)}(E)+\frac{3k}{m}\Omega_0(E),\text{ where }
m\in\mathbb{N},\,k\in\mathbb{Z};
$$
here $3k$ and $m$ are coprime numbers.

For Goryachev's solution:
$$
\Omega_{2m}^{(k)}(E)=\Omega_2^{(0)}(E)+\frac{k}{m}\Omega_0(E),
$$
where $m$ is odd, $k\in \mZ$, the numbers~$k$ and~$m$ are coprime,
while~$\Omega_0(E)=\frac{2\pi}{T(E)}$. Here, the choreography that closes
in~$mT$ corresponds to the velocity~$\Omega_m^{(k)}$ (and the choreography
that closes in~$2mT$ corresponds to the velocity~$\Omega_{2m}^{(k)}$).

\begin{figure}
\includegraphics{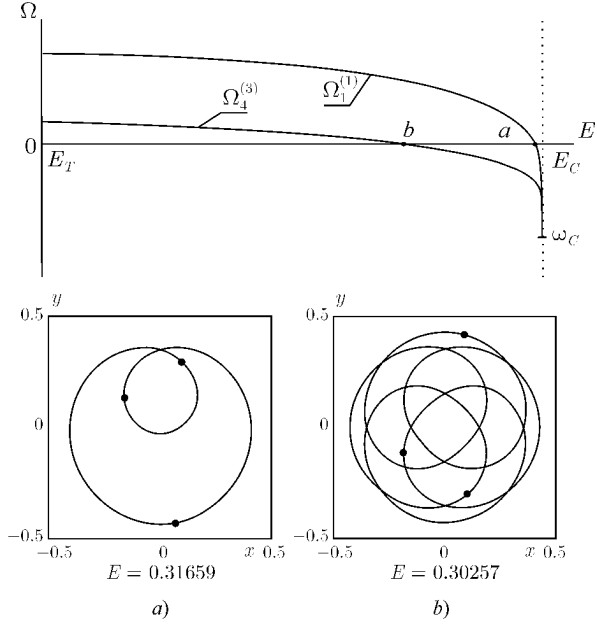}
\caption{The angular velocities of the frame
of reference for relative choreographies of three vortices; two absolute
choreographies corresponding to the points~$a$ and~$b$.}
\end{figure}
\bigskip
\begin{figure}
\includegraphics{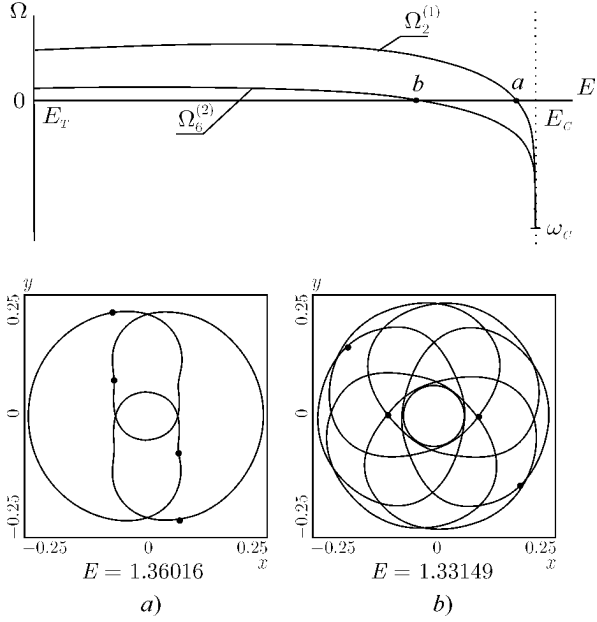}
\caption{The angular velocities of the frame
of reference for relative choreographies of four vortices; two absolute
choreographies corresponding to the points~$a$ and~$b$.}
\end{figure}

Figs.~4 and~5 show the angular velocities in
terms of the energy parameter for three and four vortices, respectively.
The roots of~$\Omega_m^{(k)}(E)=0$ correspond to the absolute
choreographies shown in Figs.~4\,a, b
and~5\,a,~b. It can be shown that there is a countable set
of energy values such that choreographies
described in Theorems~\ref{teo1} and~\ref{teo2} are absolute and simple.\vspace{-1mm}

\section{Choreographies on $\mathbb{S}^2$}

It can be shown, using the above-described methods and the reduction
described in~\cite{bmk-red}, that there are similar relative and absolute
choreographies in the systems of three and four vortices on a sphere.
Several examples of spherical choreographies of four vortices are given in
Fig.~6 (they are quite similar to those shown in
Figs.~2 and~\ref{trapez.eps}).

In~\cite{new_5} and~\cite{new_6} periodic (in a fixed frame of reference)
solutions for different numbers of vortices on a sphere are specified,
which admit different discrete symmetry groups. These motions are
disconnected choreographies. Using the group-theoretic methods, one can
construct a large number of disconnected choreographies in vortex dynamics
on a sphere, generalizing those specified in~\cite{new_5}
and~\cite{new_6}. However, the choreographies that we found cannot be
obtained using these methods.

\newpage
\begin{figure}
\includegraphics{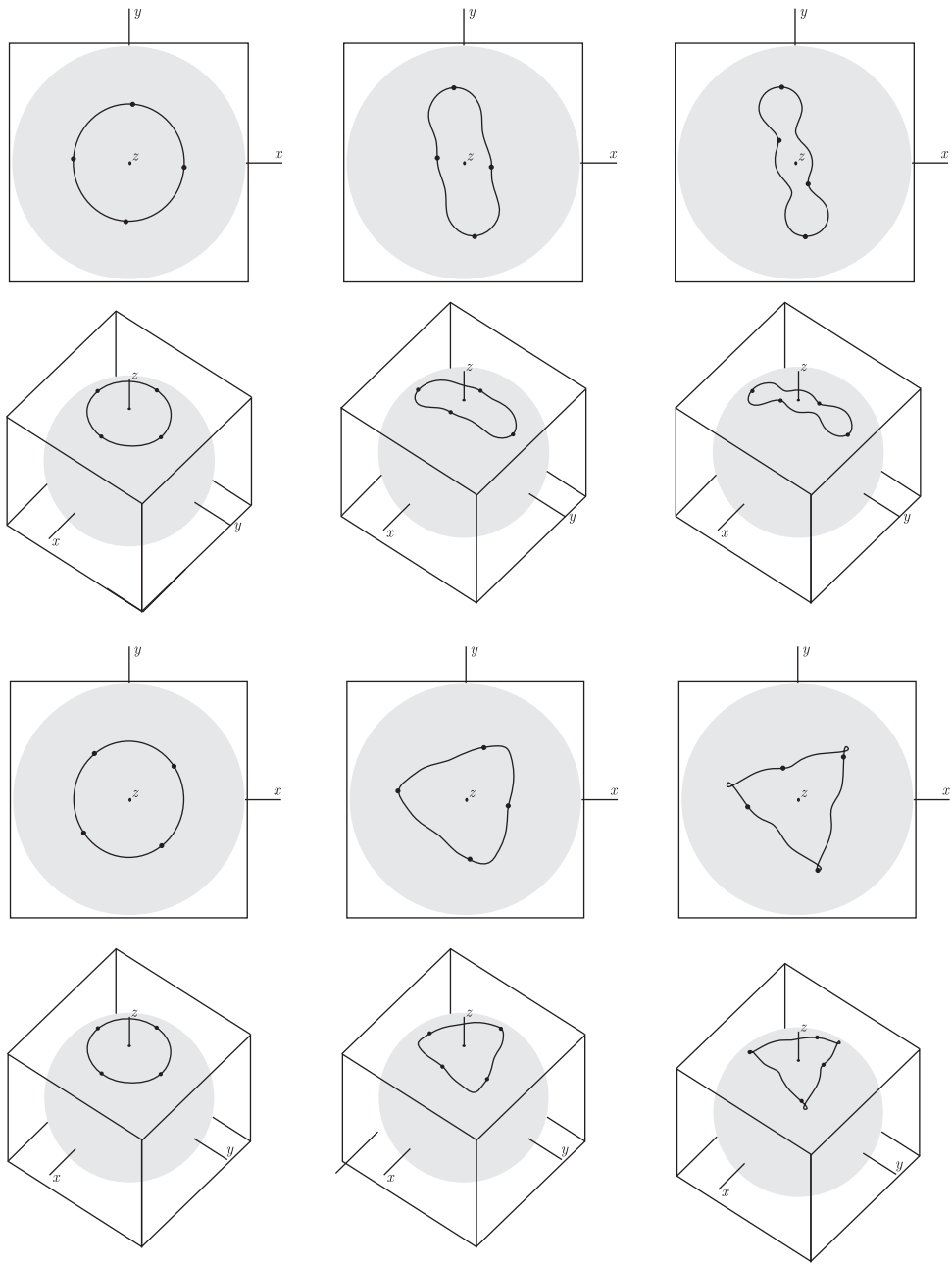}
\caption{Relative choreographies on a sphere.}
\end{figure}


\bigskip\bigskip



\newpage
\begin{thebibliography}{99}\itemsep=-1pt


\bibitem{aref1}
H.\,Aref, Point vortex motions with a center of symmetry. Phys. Fluids,
1982, v. 25 (12), p. 2183--2187.

\bibitem{aref}
H.\,Aref, N.\,Pomphrey,
{\em Integrable and chaotic motions of four vortices. I. The case of identical
vortices}. Proc. R. Soc. London, 1982, v.~380A, p.~359--387.

\bibitem{bagrets}
A.\,A.\,Bagrets, D.\,A.\,Bagrets,
{\em Nonintegrability of two problems in vortex dynamics}.
Chaos, 1997, v.~7(3), p.~368--375.

\bibitem{bogomol}
V.\,A.\,Bogomolov,
{\em Dynamics of vorticity at a sphere}. Fluid Dynamics, 1977, v.~6,
p.~863--870.

\bibitem{bogomol-2}
V.\,A.\,Bogomolov,\textit{ Model of oscillations of the atmosphere action centers.}
Izv. Atmos. Oc. Phys. 1979, v. 15 (3), p. 243--249.

\bibitem{bmk-9}
A.\,V.\,Bolsinov, A.\,V.\,Borisov, I.\,S.\,Mamaev, \emph{Lie algebras in
vortex dynamics and celestial mechanics} --- IV. Reg. \& Chaot. Dyn.
 1999, v.~4, No.~1, p.~23--50.

\bibitem{bmk-red}
A.\,V.\,Borisov, A.\,A.\,Kilin, I.\,S\,Mamaev,
\textit{Reduction and chaotic behavior of point vortices on a plane and a sphere.}
Submitted to Proceedings for the AIMS' Firth International Conference on
Dynamical Systems and Differential Equations to be published in DCDS -- B, 2005.

\bibitem{bk-dan}
A.\,V.\,Borisov, V.\,V.\,Kozlov,
\textit{Nonintegrability of the system
of interacting particles with the Dyson potential,}
Dokl. Akad. Nauk. 1999. v. 366, No 1, p.30-31 (in Russian).

\bibitem{bmlit5}
A.\,V.\,Borisov, V.\,G.\,Lebedev, \textit{Dynamics of Three Vortices on a
Plane and a Sphere -- II. General compact Case.} Reg. \& Chaot. Dyn.
 1998, v.~3, No.~2, p. 99--114.

\bibitem{bmlit8}
A.\,V.\,Borisov, I.\,S\,Mamaev, A.\,A.\,Kilin, \textit{Absolute and
relative choreographies in the problem of point vortices moving on a
plane.} Reg. \& Chaot. Dyn. (to appear).

\bibitem{new_4}
A. Chenciner, J. Gerver, R. Montgomery, C. Simo, {\it Simple choreographic
motions of $N$ bodies: a preliminary study}, Geometry, Mechanics, and
Dynamics (volume dedicated to J. Marsden, p. 287--308), Springer, 2002.

\bibitem{bmlit7}
A.\,Chenciner, P.\,Montgomery,
{\em A remarkable periodic solution of the three-body problem in the case
of equal masses}. Annals. of Math., 2000, v.~152, p.~881--901.

\bibitem{glass}
K.\,Glass,
\textit{Equilibrium configurations for a system of $N$ particles in the plane.}
Physics Letters A. 1997, v. 235, p. 591--596.

\bibitem{GoryachevSbornik}
D.\,N.\,Goryachev, {\em On certain cases of motion of rectilinear parallel
vortices\/}. Moscow, University Printing House, 1898 (in Russian).

\bibitem{grebli}
W.\,Gr\"obli,
{\em Specialle Probleme \"uber die Bewegung geradliniger paralleler
Wirbelf\"aden}.
Vierteljahrsch. d. Naturforsch. Geselsch. 1877, v.~22, p.~37--81,
p. 129--165.

\bibitem{kidambi}
R.\,Kidambi, P.\,K.\,Newton,
{\em Motion of three point vortices on a sphere}. Physica D, 1998, v.~116,
p.~143--175.

\bibitem{bmk-4}
K.\,M.\,Khanin, \emph{Quasi-periodic motions of vortex systems}. Physica
D. 1982, V.~4, p.~261-269.

\bibitem{Kurakin}
L.\,G.\,Kurakin, V.\,I.\,Yudovich, {\em The stability of stationary
rotation of a regular vortex polygon\/}. Chaos, 2002, v.~12, No.~3,
p.~574--595.

\bibitem{bmk-8}
C.\,C.\,Lim, \emph{Graph theory and special class of symplectic
transformations: the generalized Jacobi variables}. J. Math. Phys., 32(1),
1991, p.~1--7.

\bibitem{bmk-2}
P.\,K.\,Newton, \emph{The $N$-Vortex problem. Analytical Techniques},
Springer. 2001.

\bibitem{new_1}
H.\,Poincar\'{e}, {\it Th\'{e}orie des Tourbillions}. Georges
Carr\'{e}, \'{E}diteur, Paris, 1893.

\bibitem{new_5}
A. Souli\'{e}re, T. Tokieda, {\it Periodic motions of vortices on surfaces with
symmetry}, J. Fluid Mech., (2002), Vol. 460, p. 83--92.

\bibitem{new_2}
J. Synge, {\it On the motion of three vortices}, Can. J. Math. {\bf 1},
(1949), 257--270.

\bibitem{new_3}
J. Tavantzis, L. Ting, {\it The dynamics of three vortices revisited},
Phys. Fluids {\bf 31} (1988), 1392--1409.

\bibitem{Tomson}
J.\,J.\,Thomson, {\em The Corpuscular Theory of Matter}. London and
Tonbridge. 1907.

\bibitem{new_6}
T. Tokieda. {\it Tourbillions dansants}, C.R.Acad.Sci., Paris, 2001, T.
333, Ser. I, p. 943--946.

\bibitem{bmk-7}
S.\,L.\,Ziglin, \emph{Non-integrability of the problem of motion of four
point vortices\/}. Doklady AN SSSR. 1979, v.~250, No.~6, p.~1296--1300
(in Russian).


\end{thebibliography}
\end{document}